\documentclass[a4paper,11pt]{article}
\pdfoutput=1 
\usepackage{jcappub} 
\usepackage[utf8x]{inputenc}
\usepackage{multirow}
\usepackage{booktabs}
\usepackage{bm}
\usepackage{times}
\usepackage{tensor}
\title{Impacts of dark matter on the curvature of the neutron star}
\author[a,b,1]{H. C. Das,\note{Corresponding author.}}
\author[a,b]{Ankit Kumar,}
\author[c]{Bharat Kumar,}
\author[d]{S. K. Biswal,}
\author[a,b]{S. K. Patra}
\affiliation[a]{Institute of Physics, Sachivalaya Marg, Bhubaneswar-751005, India.}
\affiliation[b]{Homi Bhabha National Institute, Training School Complex, Anushakti Nagar, Mumbai 400094, India.}
\affiliation[c]{Department of Physics $\&$ Astronomy, National Institute of Technology, Rourkela, India.}
\affiliation[d]{Department of Astronomy, Xiamen University, Xiamen 361005, P. R. China.}
\emailAdd{harishdas.physics@gmail.com}
\emailAdd{ankit.k@iopb.res.in}
\emailAdd{ kumarbh@nitrkl.ac.in}
\emailAdd{subratphy@gmail.com}
\emailAdd{patra@iopb.res.in}
\abstract{
The effects of dark matter (DM) on the curvatures of the neutron star (NS) are examined by using the stiff and soft relativistic mean-field equation of states. The curvatures of the NSs are calculated with the variation of baryon density. It is found that the radial variation of different curvatures significantly affected by DM inside the star. The surface curvature is found to be more remarkable for the massive star. The effects of DM on the compactness of the maximum NS mass is less as compared to canonical star. The binding energy of the NS goes towards positive with the increase of DM momentum and makes the system unstable.}
\begin{document}
\maketitle
\flushbottom
\section{Introduction}
\label{intro}
Neutron stars (NSs) are one of the most enigmatic stellar remnants with incredibly dense core and sturdy crust, enough to hold up long-lived bulges that could produce potentially large ripples in the space, known as gravitational wave (GW). NSs can be considered as one of the best laboratory in the Universe to appraise many astrophysical models in the strong gravitational field regime \cite{Shapiro_1983,Lattimer_2007}. The observables of the NS are quite complicated to explore due to the uncertainty in the equation of state (EoS) at supra saturation density. 
    
In August 2017, the LIGO and Virgo groups detect the gravitational waves emerged from the collision of two NSs \cite{Abbott_2017,Abbott_2018}. They have considered the Universe with no dark matter (DM) in their modelling \cite{Emspak_2017}. But, the DM makes up more than 80$\%$ mass in the Universe, and it is also believed that some of the DM particles are weakly interacting, so it is customary that it will impact the NS properties up to certain extent \citep{Joglekar_2020}. It has also been reported that when a compact star rotates in the Galaxy through DM halo it captures some of the DM particles \cite{Kouvaris_2011}. The enormous gravitational force and the immense baryonic density inside the NS are responsible for the incarceration of DM Particles. The efficacy on the NS observational properties depends on the amount of DM captured by it \cite{Kouvaris_2011}. 
    
Theoretically several types of DM particles have been hypothesized and reported till date, like, weakly interacting massive particle (WIMP), feebly interacting massive particle (FIMP) etc. The WIMPs are the most abundant DM particles in the early Universe due to its freeze-out mechanism \citep{Kouvaris_2011, Ruppin_2014}. They equilibrated with the environment at freeze-out temperature and annihilate to form different standard model particles and leptons. Recently some approaches have been dedicated on the heating of NS due to deposition of kinetic energy by the DM \cite{Baryakhtar_2017,Acevedo_2019}. The annihilation of the DM particles enhance the cooling of the NSs  \citep{Kouvaris_2008, Ding_2019, Bhat_2019}. The resurrection of the DM develop its interaction with baryon which affects the structure of the NS \citep{De_Lavallaz_2010, Ciarcelluti_2011}. The accumulation of the DM inside the NS is constrained by the Chandrasekhar limit. If the accumulated quantity of the DM exceeds this limit, it can evolve into a tiny black hole and destroy the star. Different approaches have been used to calculate the NS properties with the inclusion of DM inside the NS \citep{Sandin_2009,De_Lavallaz_2010,Kouvaris_2010,Ciarcelluti_2011,Leung_2011,AngLi_2012, Panotopoulos_2017,Ellis_2018,Bhat_2019,Ivanytskyi_2019,Das_2019,Quddus_2020,Das_2020}. However, in the present scenario, we take the non-annihilating WIMP as DM candidate inside the NS. The detailed discussions can be found in our previous work \citep{Das_2020}. We observed that the addition of DM softens the EoS which results in the reduction of mass-radius of the NS.
    
According to Einstein's theory, a massive body wraps the space-time and makes curvature around it which we experience as gravity. The strength of the gravitational field at a distance $r$ from an object of mass $M$ is measured by the compactness parameter, which is defined as $\eta \equiv \frac{GM}{r c^2}$ and its value lie in between 0 and 1 \citep{Psaltis_2008, Xiao_2015}. The value of $\eta=0$ corresponds to the flat Minkowski space in special theory of relativity, while $\eta=1$ is the event horizon limit of a black hole, i.e. strongest gravitational field. Curvature is the crucial thing to quantify gravity and in determining the strength of the gravitational field \citep{Psaltis_2008}. A massive body has larger curvature than the lighter one. Some experiments had already been done to measure the curvature of the space-time. Recently, the direct detection of gravity-field curvature has also been observed by using atom interferometers \citep{Rosi_2015}. 
    
We describe the different curvature quantities briefly in the concept of general theory of relativity (GR) from the Ref. \cite{carroll_2019}. The Riemann tensor is an important quantity to measure the curvature and it has twenty independent components in four dimensions. Kretschmann scalar is defined as the square root of the full contraction of the Riemann tensor and has the same property as Riemann tensor. The Ricci tensor is the contraction of the Riemann tensor and the trace of the Ricci tensor is known as the Ricci scalar or curvature scalar. The Ricci scalar and the Ricci tensor contain all the informations about the Riemann tensor leaving only the trace-less part. Weyl tensor can be formed by removing all the contraction terms of the Riemann tensor. Physically, the Ricci scalar and the Ricci tensor measure the volumetric change of the body in presence of the tidal effect and the Weyl tensor gives information about the shape distortion of the body. However, the Riemann tensor measures both the distortion of shape and the volumetric change of the body in presence of the tidal force.
   
The strong field regime of the NS can be probed deeper with the help of modern observational instruments. The core is around 15 times denser than its surface, means most of the matter is concentrated in the core which makes the accurate measurement of $M$-$R$ profile of the NS is more complicated. But in case of the compactness and surface curvature within the star are increasing radially towards the surface. This is probably the reason that the measurement of the maximum mass-radius is more prominent for the EoS rather than the gravity. We explain the results of some recent approaches on the study of the curvature of the NS. In Ref. \cite{Kazim_2014}, it is quantified the unconstrained gravity of the NS in the framework of GR. They have calculated the curvature of the NS and noticed that GR is not well tested in the whole range of the star than EoS. Also found that the radial variation of Weyl tensor follows the power law for the large part of the star. Further, Xiao \textit{et al.} \cite{Xiao_2015} have taken both relativistic mean-field (RMF) and Skyrme-Hartree-Fock approaches and concluded that the symmetry energy affects the curvature of lighter NS significantly and have minimal effects on the massive NS. Moreover, to quantify the deviations come from GR in the strong-field regime, the detailed understanding required to study the properties of DM  in the Universe \citep{Psaltis_2008}. Therefore, in the present work, we investigate the curvature of the NS in the presence of DM in quantitative way by using RMF theory with three different parameter sets, NL3 \citep{Lalazissis_1997}, G3 \citep{Kumar_2017} and IOPB-I \citep{Kumar_2018}.
    
In the present analysis, we take the RMF formalism to calculate the EoS. The RMF model reproduces the experimental value for exotic and super-heavy nuclei \citep{Rashdan_2001, Bhuyan_2012} precisely and grant a good description of the finite nuclei up to the drip line. The extended RMF (E-RMF) formalism \citep{Kumar_2017, Kumar_2018} is in excellent agreement with all the nuclear matter (NM) properties and also satisfy most of the NS observable constraints allocated till now. In Sec. \ref{method}, we calculate the EoS of NS with the addition of DM. The curvature calculations are discussed in \ref{fcurvature}. The detailed procedures are given in \cite{Kazim_2014}. In Sec. \ref{R&D}, we present the variation of curvatures with the different quantities like baryon density, mass and radius with the addition of DM. The compactness and binding energy of the NS is also calculated with the inclusion of DM. Finally, we give a brief conclusion on the curvatures of the NS in Sec. \ref{con}.
\section{Formalism}
\label{method}
In this section, we provide the formalism required to compute the curvature of the NS in the presence of DM. First, we briefly sketch the E-RMF formalism along with DM by presenting model Lagrangian \citep{Lalazissis_1997, Kumar_2017, Kumar_2018}. All the parameters used in the E-RMF approach are fitted to reproduce the observables of finite nuclei and infinite NM. The NS matter EoS is computed in the presence of DM and hence on the NS properties to solve the Tolman-Openheimer-Volkoff (TOV) equations and its different curvatures inside on it.
\subsection{Construction of EoS using RMF approach}
\label{RMF}
The RMF Lagrangian is built from the interaction of mesons-nucleons and their self ( $\sigma^2$, $\sigma^3$, $\sigma^4$, $\omega^2$, $\omega^4$, $\rho^2$ and $\rho^4$ ) and cross-couplings ($\sigma^2-\omega^2$, $\omega^2-\rho^2$,  $\sigma-\omega^2$ and $\sigma-\rho^2$) up to fourth order. The RMF Lagrangian is discussed in these Refs.  \citep{Miller_1972,Serot_1986,Furn_1987,Reinhard_1988, Ring_1996,Frun_1997,Kumar_2017,Kumar_2018}. The RMF Lagrangian for NM system is  
\begin{eqnarray}
{\cal L}_{nucl.} & = &  \sum_{\alpha=p,n} \bar\psi_{\alpha}
\Bigg\{\gamma_{\mu}\bigg(i\partial^{\mu}-g_{\omega}\omega^{\mu}-\frac{1}{2}g_{\rho}\vec{\tau}_{\alpha}\!\cdot\!\vec{\rho}^{\,\mu}\bigg)-\bigg(M_{nucl.} 
-g_{\sigma}\sigma-g_{\delta}\vec{\tau}_{\alpha}\!\cdot\!\vec{\delta}\bigg)\Bigg\} \psi_{\alpha}+\frac{1}{2}\partial^{\mu}\sigma\,\partial_{\mu}\sigma
\nonumber \\
&&  
-\frac{1}{2}m_{\sigma}^{2}\sigma^2+\frac{\zeta_0}{4!}g_\omega^2(\omega^{\mu}\omega_{\mu})^2-\frac{\kappa_3}{3!}\frac{g_{\sigma}m_{\sigma}^2\sigma^3}{M_{nucl.}}-\frac{\kappa_4}{4!}\frac{g_{\sigma}^2m_{\sigma}^2\sigma^4}{M_{nucl.}^2}
+\frac{1}{2}m_{\omega}^{2}\omega^{\mu}\omega_{\mu}-\frac{1}{4}W^{\mu\nu}W_{\mu\nu}
\nonumber\\
&&
+\frac{\eta_1}{2}\frac{g_{\sigma}\sigma}{M_{nucl.}}m_\omega^2\omega^{\mu}\omega_{\mu}+\frac{\eta_2}{4}\frac{g_{\sigma}^2\sigma^2}{M_{nucl.}^2}m_\omega^2\omega^{\mu}\omega_{\mu}+\frac{\eta_{\rho}}{2}\frac{m_{\rho}^2}{M_{nucl.}}g_{\sigma}\sigma\bigg(\vec\rho^{\,\mu}\!\cdot\!\vec\rho_{\mu}\bigg)+\frac{1}{2}m_{\rho}^{2}\bigg(\vec\rho^{\mu}\!\cdot\!\vec\rho_{\mu}\bigg)
\nonumber\\
&&
-\frac{1}{4}\vec R^{\mu\nu}\!\cdot\!\vec R_{\mu\nu}
-\Lambda_{\omega}g_{\omega}^2g_{\rho}^2\big(\omega^{\mu}\omega_{\mu}\big)
\big(\vec\rho^{\,\mu}\!\cdot\!\vec\rho_{\mu}\big)+\frac{1}{2}\partial^{\mu}\vec\delta\,\partial_{\mu}\vec\delta-\frac{1}{2}m_{\delta}^{2}\vec\delta^{\,2},
\label{lag}
\end{eqnarray}
where $M_{nucl}$ (= 939 MeV) is the mass of the nucleon. $m_\sigma$, $m_\omega$, $m_\rho$ and $m_\delta$ are the masses and $g_\sigma$, $g_\omega$, $g_\rho$ and $g_\delta$ are the coupling constants for the $\sigma$, $\omega$, $\rho$ and $\delta$ mesons respectively. $\kappa_3$ (or $\kappa_4$) and $\zeta_0$ are the self-interacting coupling constants of the $\sigma$ and $\omega$ mesons respectively. $\eta_1$, $\eta_2$, $\eta_\rho$ and $\Lambda_\omega$ are the coupling constants of non-linear cross-coupled terms. The quantities $W^{\mu\nu}$ and $\vec R^{\mu\nu}$ being the field strength tensors for the $\omega$ and $\rho$ mesons respectively, defined as $W^{\mu\nu}$ = $\partial^\mu\omega^\nu-\partial^\nu\omega^\mu$ and $\vec R^{\mu\nu}$ = $\partial^\mu\vec\rho^{\,\nu}-\partial^\nu\vec\rho^{\,\mu}$. The $\vec\tau$ are the Pauli matrices and behave as the isospin operator. Parameters and saturation properties for NL3 \citep{Lalazissis_1997}, G3 \citep{Kumar_2017} and IOPB-I \citep{Kumar_2018} along with the empirical/experimental values are given in Table \ref{table1}.

The meson fields for the NM system are calculated by solving the mean-field equation of motions \citep{Kumar_2017, Kumar_2018, Das_2019} in a self-consistent way. The energy density (${\cal{E}}_{nucl.}$) and pressure ( $P_{nucl.}$) are calculated using the energy-momentum stress-tensor technique which is given by \citep{Walecka_74, NKGb_1997}
\vspace{5cm}
\begin{eqnarray}
{\cal{E}}_{nucl.}&=& \frac{\gamma}{(2\pi)^{3}}\sum_{i=p,n}\int_0^{k_i} d^{3}k E_{i}^\star (k_i)+\rho_bW +\frac{1}{2}\rho_{3}R
+\frac{ m_{s}^2\Phi^{2}}{g_{s}^2}\Bigg(\frac{1}{2}+\frac{\kappa_{3}}{3!}
\frac{\Phi }{M_{nucl.}} + \frac{\kappa_4}{4!}\frac{\Phi^2}{M_{nucl.}^2}\Bigg)
\nonumber\\
&&
-\frac{1}{4!}\frac{\zeta_{0}W^{4}}
{g_{\omega}^2}-\frac{1}{2}m_{\omega}^2\frac{W^{2}}{g_{\omega}^2}\Bigg(1+\eta_{1}\frac{\Phi}{M_{nucl.}}+\frac{\eta_{2}}{2}\frac{\Phi ^2}{M_{nucl.}^2}\Bigg)
-\Lambda_{\omega}(R^{2}\times W^{2})-\frac{1}{2}\Bigg(1+\frac{\eta_{\rho}\Phi}{M_{nucl.}}\Bigg)
\nonumber\\ 
&&
\times\frac{m_{\rho}^2}{g_{\rho}^2}R^{2}+\frac{1}{2}\frac{m_{\delta}^2}{g_{\delta}^{2}}D^{2},
\label{e_nm}
\end{eqnarray}
\noindent
\begin{eqnarray}
P_{nucl.} &=&  \frac{\gamma}{3 (2\pi)^{3}}\sum_{i=p,n}\int_0^{k_i} d^{3}k \frac{k^2}{E_{i}^\star (k_i)}+\frac{1}{4!}\frac{\zeta_{0}W^{4}}{g_{\omega}^2}
-\frac{ m_{s}^2\Phi^{2}}{g_{s}^2}\Bigg(\frac{1}{2}+\frac{\kappa_{3}}{3!}
\frac{\Phi }{M_{nucl.}}+ \frac{\kappa_4}{4!}\frac{\Phi^2}{M_{nucl.}^2}\Bigg)
\nonumber\\
&&
+\frac{1}{2}m_{\omega}^2\frac{W^{2}}{g_{\omega}^2}\Bigg(1+\eta_{1}\frac{\Phi}{M_{nucl.}}+\frac{\eta_{2}}{2}\frac{\Phi ^2}{M_{nucl.}^2}\Bigg)
+\Lambda_{\omega} (R^{2}\times W^{2})+\frac{1}{2}\Bigg(1+\frac{\eta_{\rho}\Phi}{M_{nucl.}}\Bigg)\frac{m_{\rho}^2}{g_{\rho}^2}R^{2}
\nonumber\\
&&
-\frac{1}{2}\frac{m_{\delta}^2}{g_{\delta}^{2}}D^{2} \nonumber.\\
\label{p_nm}
\end{eqnarray}
Where $\Phi$, $W$, $R$ and $D$ are the redefined fields for $\sigma$, $\omega$, $\rho$ and $\delta$ mesons as
$\Phi = g_s\sigma^0$, $W = g_\omega \omega^0$, $R$ = g$_\rho\vec{\rho}$ $^0$ and $D=g_\delta\delta^0$ respectively. The $E_{i}^\star(k_i)$=$\sqrt {k_i^2+{M_{i}^\star}^2}$, where $M_i^\star$ is the effective mass and $k_i$ is the momentum of the nucleon and $\gamma$ is the spin degeneracy factor which is equal to 2 for individual nucleons.
\begin{table}
\caption{The parameter sets NL3 \citep{Lalazissis_1997}, G3 \citep{Kumar_2017} and  IOPB-I \citep{Kumar_2018} are listed. All the coupling constants are dimensionless. The NM parameters are given in the lower panel including with empirical/experimental values at the saturation density. The references are $[a]$,$[b]$, $[c]$ $\&$ $[d]$ \citep{Zyla_2020}, $[e] \& [f]$ \citep{Bethe_1971}, $[g]$ \citep{Garg_2018}, $[h] \& [i]$ \citep{Danielewicz_2014}, and $[j]$ \citep{Zimmerman_2020}.}
\centering
\begin{tabular}{cccccccccc}
\hline
\hline
\multicolumn{1}{c}{Parameter}
&\multicolumn{1}{c}{NL3}
&\multicolumn{1}{c}{G3}
&\multicolumn{1}{c}{IOPB-I}
&\multicolumn{1}{c}{Empirical/Expt. Value}\\
\hline
$m_{\sigma}/M_{nucl.}$  &  0.541  &  0.559&0.533 & 0.426 -- 0.745 $[a]$\\
$m_{\omega}/M_{nucl.}$  &  0.833 &  0.832&0.833 & 0.833 -- 0.834 $[b]$ \\
$m_{\rho}/M_{nucl.}$  &  0.812 &  0.820&0.812 & 0.825 -- 0.826 $[c]$\\
$m_{\delta}/M_{nucl.}$   & 0.0  &   1.043&0.0 & 1.022 -- 1.064 $[d]$\\
$g_{\sigma}/4 \pi$  &  0.813   &  0.782 &0.827 & \\
$g_{\omega}/4 \pi$  &  1.024  &  0.923&1.062 & \\
$g_{\rho}/4 \pi$  &  0.712 &  0.962 &0.885 & \\
$g_{\delta}/4 \pi$  &  0.0  &  0.160& 0.0 &\\
$k_{3} $   &  1.465  &    2.606 &1.496 &\\
$k_{4}$  &  -5.688  & 1.694 &-2.932  &\\
$\zeta_{0}$  &  0.0  &  1.010  &3.103 & \\
$\eta_{1}$  &  0.0  &  0.424 &0.0  &\\
$\eta_{2}$  &  0.0  &  0.114 &0.0 & \\
$\eta_{\rho}$  &  0.0 &  0.645& 0.0  &\\
$\Lambda_{\omega}$  &  0.0 &  0.038&0.024 &\\ \hline
$\rho_0$ ($fm^{-3}$) & 0.148 & 0.148 &0.149& 0.148 -- 0.185 \ $[e]$ \\
$BE (MeV)$ & -16.29 & -16.02 &-16.10& -15.00 -- 17.00 \ $[f]$\\
$K (MeV)$ & 271.38 & 243.96&222.65 & 220 -- 260 \qquad $[g]$\\
$J (MeV)$ & 37.43 & 31.84&33.30 & 30.20 -- 33.70\quad $[h]$\\
$L (MeV)$ & 118.65 & 49.31 & 63.58& 35.00 -- 70.00 \quad$[i]$\\
$K_{sym} (MeV)$ & 101.34 & -106.07 & -37.09 & -174 -- -31 \qquad $[j]$ \\
$Q_{sym}$ (MeV) & 177.90 & 915.47&862.70 & -----------\\
\hline
\hline
\end{tabular}
\label{table1}
\end{table}

    Inside the NS, many particles like hyperons, nucleons and leptons are present. The neutron decays to proton, electron and anti-neutrino inside the NS \citep{NKGb_1997}. This process is called as $\beta$--decay. To maintain the charge neutrality condition, the inverse $\beta$--decay process occurred. The process can be expressed as
    \begin{eqnarray}
    n \rightarrow p+e^-+\bar\nu, \nonumber\\ 
    p+e^-\rightarrow n+\nu .
    \label{betaeq}
    \end{eqnarray}
    To maintain the stability of NSs, there must have both $\beta$-equilibrium and charge-neutrality conditions, which are expressed in term of chemical potential
    \begin{eqnarray}
    \mu_n &=& \mu_p +\mu_e,   \nonumber \\
    \mu_e &=& \mu_\mu,\nonumber\\
    \rho_p &=& \rho_e +\rho_\mu.
    \label{betaandcharge}
    \end{eqnarray}
    Where $\mu_n$, $\mu_p$, $\mu_e$, and $\mu_\mu$ are the chemical potentials of neutrons, protons, electrons, and muons, respectively. When the chemical potential of electron is equal to the muon rest mass, then muon appear inside the NS.
    The chemical potentials $\mu_n$, $\mu_p$, $\mu_e$, and $\mu_\mu$ are given by \citep{Das_2020}
    \begin{equation}
    \mu_{n,p} = g_\omega \omega_0 \pm g_\rho \rho_0+\sqrt{k_{n,p}^2+ (M_{n,p}^\star)^2},
    \label{eqnp}
    \end{equation} 
    \begin{equation}
    \mu_{e,\mu} = \sqrt{k_{e,\mu}^2+ m_{e,\mu}^2},
    \label{eqemu}   
    \end{equation}
    where $M_n^\star$ and $M_p^\star$ are the effective masses of neutron and proton respectively. The particle fraction inside the NS is calculated by solving Eqs. (\ref{betaandcharge}) using the Eqs. (\ref{eqnp} -- \ref{eqemu}) for a given baryon density by self-consistently. The energy density and pressure of NS are given by,
    \begin{eqnarray}
    {\cal {E}}_{NS} &=& {\cal {E}}_{nucl.}+ {\cal {E}}_{l}, \nonumber \\ and \hspace{1cm}
    P_{NS}&=& P_{nucl.}+ P_l,
    \label{eqNS}
    \end{eqnarray}
    with, 
    \begin{equation}
    {\cal {E}}_{l} = \sum_{l=e,\mu}\frac{2}{(2\pi)^{3}}\int_0^{k_l} d^{3}k \sqrt{k^2 + m_l^2 },
    \label{eqel}
    \end{equation}
    and
    \begin{equation}
    P_{l} = \sum_{l=e,\mu}\frac{2}{3(2\pi)^{3}}\int_0^{k_l} \frac{d^{3}k \hspace{0.2cm}k^2} {\sqrt{k^2 + m_l^2}}.
    \label{eqpl}
    \end{equation}
    Where ${\cal{E}}_{l}$, $P_{l}$ and $k_l$ are the energy density, pressure and Fermi momentum for leptons respectively. The Eq. (\ref{eqNS}) gives the total energy and pressure of the NS. 
    \subsection{Interaction of DM candidates in NS}\label{NDM}
    DM particles accreted inside the NS core due to its high gravitational field \citep{Goldman_1989,Kouvaris_2008,Xiang_2014,Das_2019} and the amount of accreted DM depends directly on its evolving life time. In this scenario, we consider the Neutralino \citep{Martin_1998,Panotopoulos_2017,Das_2019,Das_2020} as a fermionic DM candidate which interacts with nucleon via SM Higgs. The detailed formalism has been taken from our previous analysis \citep{Das_2020} and the total Lagrangian is written as:
    \begin{eqnarray}
    {\cal{L}}_{tot.} & = & {\cal{L}}_{NS} + \bar \chi \left[ i \gamma^\mu \partial_\mu - M_\chi + y h \right] \chi +  \frac{1}{2}\partial_\mu h \partial^\mu h 
    - \frac{1}{2} M_h^2 h^2 + f \frac{M_{nucl.}}{v} \bar \varphi h \varphi , 
    \label{eqdm}
    \end{eqnarray}
    where ${\cal{L}}_{NS}$ is the NS Lagrangian and $\varphi$ and $\chi$ are the nucleonic and DM wave functions respectively. $h$ is the Higgs field. The values of the parameters like $y (=0.07)$, $f (=0.35)$ and $v (=246$ GeV) are given in \citep{Das_2020}. From the Lagrangian (Eq. (\ref{eqdm}), we get the total energy density (${\cal{E}}_{tot.}$) and pressure ($P_{tot.}$) for NS with DM given as \citep{Das_2020} 
    \begin{eqnarray}
    {\cal{E}}_{tot.}& = &  {\cal{E}}_{NS} + \frac{2}{(2\pi)^{3}}\int_0^{k_f^{DM}} d^{3}k \sqrt{k^2 + (M_\chi^\star)^2 } 
    + \frac{1}{2}M_h^2 h_0^2 ,
    \label{etot}
    \end{eqnarray}
    and
    \begin{eqnarray}
    P_{tot.}& = &  P_{NS} + \frac{2}{3(2\pi)^{3}}\int_0^{k_f^{DM}} \frac{d^{3}k \hspace{1mm}k^2} {\sqrt{k^2 + (M_\chi^\star)^2}} 
    - \frac{1}{2}M_h^2 h_0^2 ,
    \label{ptot}
    \end{eqnarray} 
    where $k_f^{DM}$ is the DM Fermi momentum. The last term of Eqs. (\ref{etot}) and (\ref{ptot}) represents the potential of the Higgs field, where $M_h$ is the mass of the Higgs which is 125 GeV and $h_0$ is the Higgs field calculated by applying the mean-field approximation ( see Eq. (8c) of \cite{Das_2019}). We find that the Higgs field contribution to both energy density and pressure are very small ($10^{-12}$ MeV). Since the interaction between nucleon and Higgs is very small, the vacuum expectation value $v$ (= 246 GeV) don't change inside the neutron star. The $M_{n,p}^{\star}$ and $M_{\chi}^{\star}$ are the effective masses of nucleon and DM, which are given as
    \begin{eqnarray}
    M_{n,p}^\star &=& M_{nucl.}+ g_\sigma \sigma_0 \mp g_\delta \delta_0 - \frac{f M_{nucl.}}{v}h_0, 
    \nonumber\\
    M_\chi^\star &=& M_\chi -y h_0,
    \label{effndm}
    \end{eqnarray}
    where the $\sigma_0$ and $\delta_0$ are the mean-field for $\sigma$ and $\delta$ mesons respectively.
    \subsection{Experimental evidences}
    In the present formalism, two coupling constants has major significance (i) DM-Higgs coupling ($y$), and (ii) nucleon-Higgs form factor ($f$). The detail prescription given as follows \\
    (i) The direct detection experiment don't show any collision events till now but they gave an upper bounds on the WIMP-nucleon scattering cross-section which is function of the DM mass. The WIMP undergoes elastic collision with the detector nucleus ( in quark level) by the Higgs exchange. Therefore the interaction Lagrangian which contains both DM wave function ($\chi$) and quark wave function ($q$) can be written as \cite{Bhat_2019}
    \begin{equation}
    {\cal{L}}_{int}=\alpha_q \Bar{\chi}\chi\Bar{q}q,
    \end{equation}
    where $\alpha_q=\frac{y f m_q}{vM_h^2}$. $q$ is the valence quark, $f$ is nucleon-Higgs form factor and $m_q$ is the mass of the quark. In this calculations, the values of $y$ and $f$ are taken as 0.07 and 0.35. The spin independent cross-section for the fermionic dark matter can be written as \cite{Bhat_2019}
    \begin{equation}
    \sigma_{SI}=\frac{y^2f^2M_n^2}{4\pi}\frac{\mu_r}{v^2M_h^2} ,
    \end{equation}
    where $M_n$ (= 939 MeV) is the nucleon mass and $\mu_r$ is the reduced mass $\frac{M_nM_\chi}{M_n+M_\chi}$, $M_\chi$ is the mass of the DM particle. We calculated the $\sigma_{SI}$ for three different masses of dark matter 50, 100, and 200 GeV and their corresponding cross-section found to be 9.43, 9.60 and 9.70 of the order of ($10^{-46}$ cm$^2$) respectively. That means the predicted values are consistent with the direct detection experiment like XENON-1T \cite{Xenon1T_2016}, PandaX-II \cite{PandaX_2016} and LUX \cite{LUX_2017} with 90\% confidence level. In case of LHC, which produced various WIMP-nucleon cross section limit in the range from $10^{-40}$ to $10^{−50}$ cm$^2$ depending on the dark matter production models \cite{PandaX_2016}. Thus our model also satisfies the LHC limit. Therefore in the present calculations, we constrained the value of $y$ from both the direct detection experiments and the LHC results.\\
    (ii) Nucleon-Higgs form factor ($f$) had been calculated in Ref. \cite{Djouadi_2012} using the implication of both lattice QCD \cite{Czarnecki_2010} and MILC results \cite{MILC_2009} whose value is $0.33_{-0.07}^{+0.30}$ \cite{Aad_2015}. The taken value of $f$ (= 0.35 ) in this calculation lies in the region.
    \subsection{Mass and Radius of the NS}
    Here we calculate the NS observables like $M$ and $R$ etc. using TOV equations. Hence, we take the EoSs of NS with DM and input to the TOV equations \citep{TOV1,TOV2} are given as
    \begin{eqnarray}
    \frac{dP_{tot.}(r)}{dr}= - \frac{(P_{tot.}(r)+{\cal{E}}_{tot.}(r))(m(r)+4\pi r^3 P_{tot.}(r))}{r(r-2m(r))}, \nonumber\\
    \label{TOV1}
    \end{eqnarray}
    and 
    \begin{eqnarray}
    \frac{dm(r)}{dr}=4\pi r^2 {\cal{E}}_{tot.}(r),
    \label{TOV2}
    \end{eqnarray}
    where ${\cal{E}}_{tot.}(r)$ and $P_{tot.}(r)$ are the total energy and pressure density as a function of radial distance. $m(r)$ is the gravitational mass, and $r$ is the radial parameter. These two coupled equations are solved to get the mass and radius of the NS at certain central density.
    \subsection{Mathematical formulation for different curvatures}
    \label{fcurvature}
    We adopt the mathematical form of different curvature quantities from \cite{Kazim_2014}, which measure the curvatures for both inside and outside the star. The curvatures are Ricci scalar, Ricci tensor, Riemann tensor and Weyl tensor, which are formulated as 
    
    The Ricci scalar
    \begin{equation}
    {\cal R}(r)=8\pi\bigg[{\cal{E}}_{tot.}(r) -3 P_{tot.}(r)\bigg],
    \label{RS}
    \end{equation}
    the full contraction of the Ricci tensor
    \begin{equation}
    {\cal J}(r) \equiv \sqrt{{\cal R}_{\mu \nu} {\cal R}^{\mu \nu}} = \bigg[(8\pi)^2 \left[ {\cal{E}}_{tot.}^2(r) + 3P_{tot.}^2(r)\right]\bigg]^{1/2},
    \label{RT}
    \end{equation}
    the Kretschmann scalar (full contraction of the Riemann tensor)
    \begin{eqnarray}
    {\cal{K}}(r)&\equiv&\sqrt{{\cal{R}}^{\mu\nu\rho\sigma}{\cal{R}}_{\mu\nu\rho\sigma}}
    \nonumber\\
    &&
    =\bigg[(8\pi)^2[3{\cal{E}}_{tot.}^2(r)+3P_{tot.}^2(r)
    +2P_{tot.}(r){\cal{E}}_{tot.}(r)]
    -\frac{128{\cal{E}}_{tot.}(r)m(r)}{r^3}+\frac{48m^2(r)}{r^6}\bigg]^{1/2},
    \label{KS}
    \end{eqnarray}
    and the full contraction of the Weyl tensor
    \begin{equation}
    {\cal W}(r) \equiv \sqrt{{\cal C}^{\mu \nu \rho \sigma }{\cal C}_{\mu \nu \rho \sigma}} = \bigg[\frac43 \left( \frac{6m(r)}{r^3} - 8\pi {\cal{E}}_{tot.}(r) \right)^2\bigg]^{1/2}.
    \label{WT}
    \end{equation}
    Where  ${\cal{E}}_{tot.}$, $P_{tot.}$, $m(r)$ and $r$ are the energy density, pressure, mass and radius of the NS respectively. At  the surface $m\rightarrow M$ due to $r \rightarrow R$. The Ricci tensor and Ricci scalar vanish outside the star because they depends on the ${\cal{E}}_{tot.}(r)$, $P_{tot.}(r)$ which are zero outside the star. But, there is a non-vanishing component of the Riemann tensor which does not vanish; $\tensor{{\cal R}}{^1_{010}}=-\frac{2M}{R^3}=- \xi$, even in the outside of the star \citep{Kazim_2014, Xiao_2015}. So the Riemann tensor is the more relevant quantity to measure the curvature of the stars than others. Kretschmann scalar is the square root of the full contraction of the Riemann tensor. The vacuum value for both $\cal{K}$ and $\cal{W}$ is $\frac{4\sqrt{3 M}}{R^3}$ as easily can see Eqs. (\ref{KS}) and (\ref{WT}). There-fore, one can take $\cal K$ and $\cal W$ as two reasonable measures the curvature within the star. 
    \section{Results and Discussion}
    \label{R&D}
    \subsection{Choice of parameter sets}
    In the present work, three sets of parameter are chosen, namely NL3 \citep{Lalazissis_1997}, IOPB-I \citep{Kumar_2018} and G3 \citep{Kumar_2017}. The NL3 parameter set corresponds to the standard RMF model, which contains non-linear interactions (self-interaction of the sigma mesons). It has relatively high incompressibility $K$ in comparison to the other two parameter sets (see Table \ref{table1}). The NM properties  and their empirical/experimental values of all the three-parameter sets are given in Table \ref{table1}. In case of IOPB-I, there are two extra coupling parameters, $\Lambda_\omega$ and $\zeta_0$ on top of the NL3 set. These two coupling parameters play a vital role in both finite and infinite NM system \citep{Singh_2014,Biswal_2015,Kumar_2018}. The parameter $\Lambda_\omega$ controls the symmetry energy (or neutron skin thickness of finite nuclei) as well as the maximum mass of the NS \citep{Horowitz_2001}. $\zeta_0$, which is the coupling constants for the self-interaction of the vector-meson, affects the EoS at higher density \citep{Toki_1994,Muller_1996}. So it is imperative to include IOPB-I parameter set for the study of the NS physics. G3 interaction holds all the coupling parameters present in the Lagrangian discussed in Sec. \ref{RMF}. One can note that G3 parameter has six extra couplings constants ($\eta_1$, $\eta_2$, $\eta_{\rho}$, $g_{\delta}$, $\Lambda_{\omega}$ and $\zeta_0$) in comparison to the NL3 parameter set. All the couplings constants of the Lagrangian in Eq. (\ref{lag}) are obtained by fitting the several properties of finite nuclei and infinite NM at saturation density. The NL3 being the stiffest EoS enrich with the higher NS mass (2.774 $M_{\odot}$) in comparison to other parameter sets. The maximum mass, canonical radius and tidal deformability are calculated by using IOPB-I parameter set are in excellent agreement with the data suggested by GW170817 observational event \citep{Abbott_2017, Abbott_2018}. We give the results with these parameter sets, for the comparative study and better understanding of the parametric dependency of the DM effects on the curvatures of NS. 
    \subsection{Equation of State}
    \label{forEoS}
\begin{figure}
      \centering
        \includegraphics[width=0.6\columnwidth]{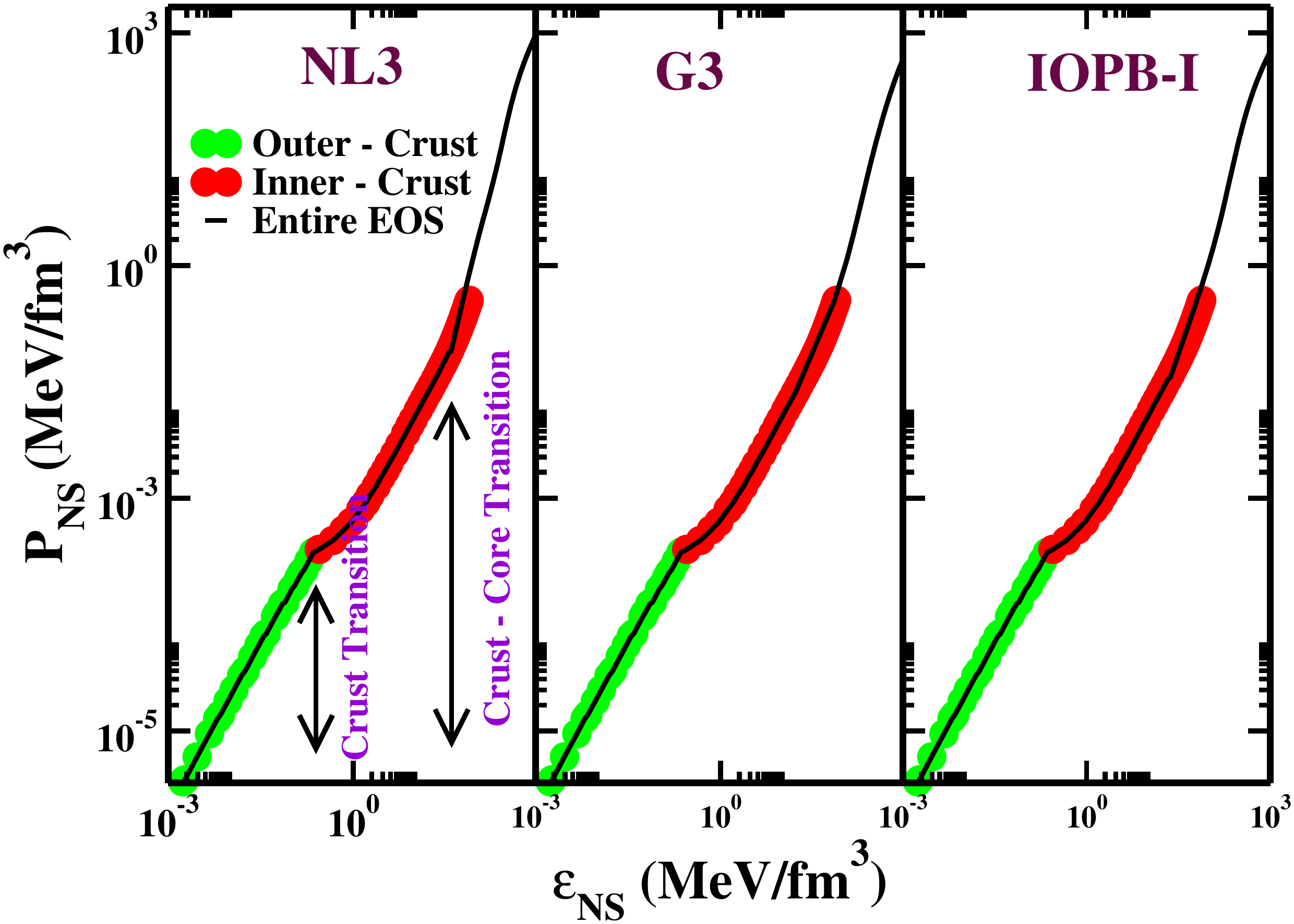}
        \caption{ (colour online) The EoS of the NS is shown for different parameter sets for core part ($<5\rho_0$), where $\rho_0$ is the NM saturation density. The red and green shaded line represents the inner crust ( 3$\times10^{-4}$ -- 8$\times10^{-2}$ fm$^{-3}$) and outer crust (6$\times10^{-12}$ -- 2.61$\times10^{-4}$ fm$^{-3}$) of Ref. \citep{BKS_2015} respectively.}
        \label{EoS}
    \end{figure}

    EoS is the most vital equipment to understand the properties of NS. We use EoS emerged from different parameter sets as mentioned above, NL3 \citep{Lalazissis_1997}, G3 \citep{Kumar_2017} and IOPB-I \citep{Kumar_2018} to explore the curvature of NS. The calculated EoS in Eqs. (\ref{eqNS}) is the only for the core part of the NS and it is different for NL3, G3 and IOPB-I parameter sets. However, for the crust part (both inner and outer crust), we adopt the Sharma et. al. EoS \citep{BKS_2015} which is added to form unified EoS for the whole density range. The entire EoS is depicted in Fig. \ref{EoS}. Moreover, we have found in our recent work \cite{Das_2020} that the EoS becomes softer with the addition of DM and reduces the $M$ and $R$ of the NS. The maximum mass, radii and its corresponding central densities are given in Table \ref{table2} for different DM momenta.
    \subsection{Contribution to curvatures by various parameters}
\begin{figure}[tbp]
       \centering
       \includegraphics[width=0.6\textwidth]{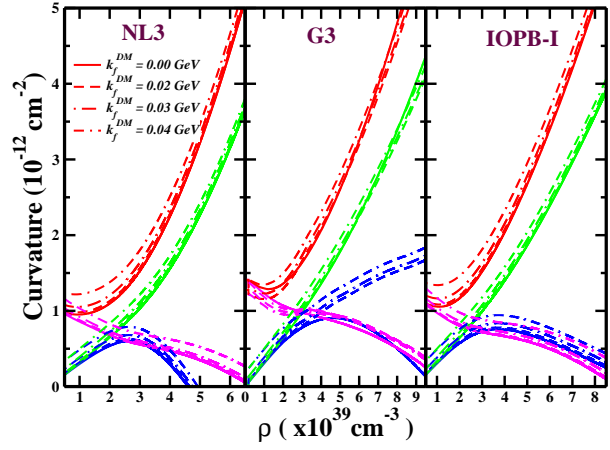}
       \caption{(colour online) The variation of different curvatures  $\cal{K}$ (red), $\cal{J}$ (green), $\cal{R}$ (blue) and $\cal{W}$ (magenta) with baryon density for NL3 (left), G3 (middle) and IOPB-I (right) in the presence of DM for corresponding maximum mass.}
        \label{density}
    \end{figure}

    We calculate various curvatures like $\cal{K}, \cal{J}, \cal{R}$ and $\cal{W}$ of the NS in the presence of DM. The curvature of the NS with the variation of baryon density with different DM momentum for all the assumed parameter sets are shown in Fig. \ref{density}. At the low density region (near to the surface), the curvatures $\cal{J}$ and $\cal{R}$ almost vanish due to their zero vacuum value (see Eq. \ref{RS} and \ref{RT}) but the curvatures $\cal{K}$ and $\cal{W}$ approach each other at the local maximum $\frac{4\sqrt{3} M}{R^3}$. In the high dense portion the $\cal K$ and $\cal J$ procure larger curvature than others. All the curvatures increase with the $k_f^{DM}$ as shown in Fig. \ref{density}. This is due to the fact that the EoS becomes more softer with $k_f^{DM}$, which gives more curvature as compared to stiffer EoS. G3 gives large curvature as compare to IOPB-I and NL3 due to its soft EoS. 
    
\begin{figure}[tbp]
        \centering
        \includegraphics[width=0.7\textwidth]{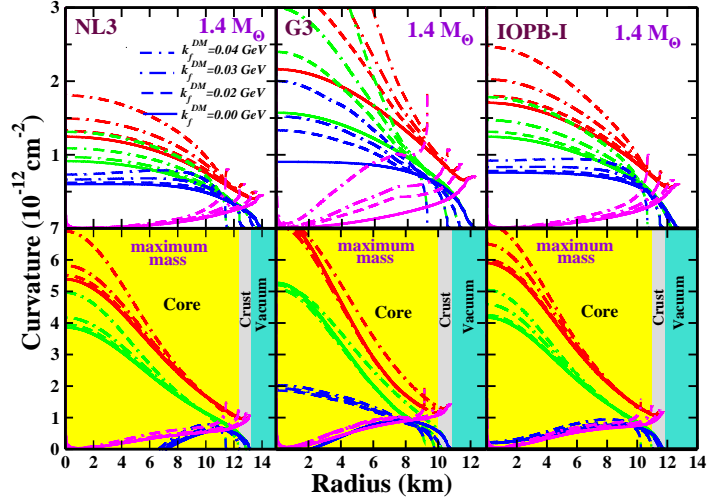}
        \caption{(colour online) The radial variation of all the curvatures  $\cal{K}$ (red), $\cal{J}$ (green), $\cal{R}$ (blue) and $\cal{W}$ (magenta) for NL3 (left), G3 (middle) and IOPB-I (right) in the presence of DM. The yellow, grey and cyan colour regions represents the core, the crust and the vacuum of the NS respectively. The curvatures $\cal R$ become negative for maximum mass star without DM. This negative value is about 6.5 km, 2 km and 1.2 km for NL3, G3 and IOPB-I respectively. This is due to the limitations of EoSs obtained from the assumed parameter sets which are not fully compatible with special relativity. The EoS violates ${\cal{E}}\geq3P$ in higher densities limit \cite{Kazim_2014}.}
        \label{curv1}
    \end{figure}

    The radial variation of the curvatures with the addition of DM is shown in Fig \ref{curv1}.  All the curvatures are maximum at the centre of the star except Weyl tensor. However, the Ricci scalar is negative within the star (for maximum mass), as shown in Fig. \ref{curv1}. At the surface of the star, ${\cal{E}}_{tot.}=0$ and $P_{tot.}=0$, so $\cal{K}$ and $\cal{W}$ are equal (see Eqs. (\ref{KS}) and (\ref{WT})). Near the surface of the NS, the $\cal{J}$ and $\cal{R}$ approach to zero. If we assume that the NS has uniform density, i.e. $m=\frac{4}{3}\pi r^3\rho$, then the Eq. (\ref{WT}) is equal to zero. Therefore, the $\cal{W}$ tends to zero at the core. As we go from outer crust to the surface, the density in this region is like diffuse state, so that $\cal{W}$ is maximum at the surface. Thus, it can be concluded that $\cal K$ and $\cal W$ attain different values within the star and approach each other at the crust and give identical value in vacuum as shown in Fig. \ref{curv1}. The radial variation of curvatures follow the same trends in presence of the DM, but the magnitude of the curvature is more. 
    
\begin{figure}[tbp]
        \centering
        \includegraphics[width=0.6\columnwidth]{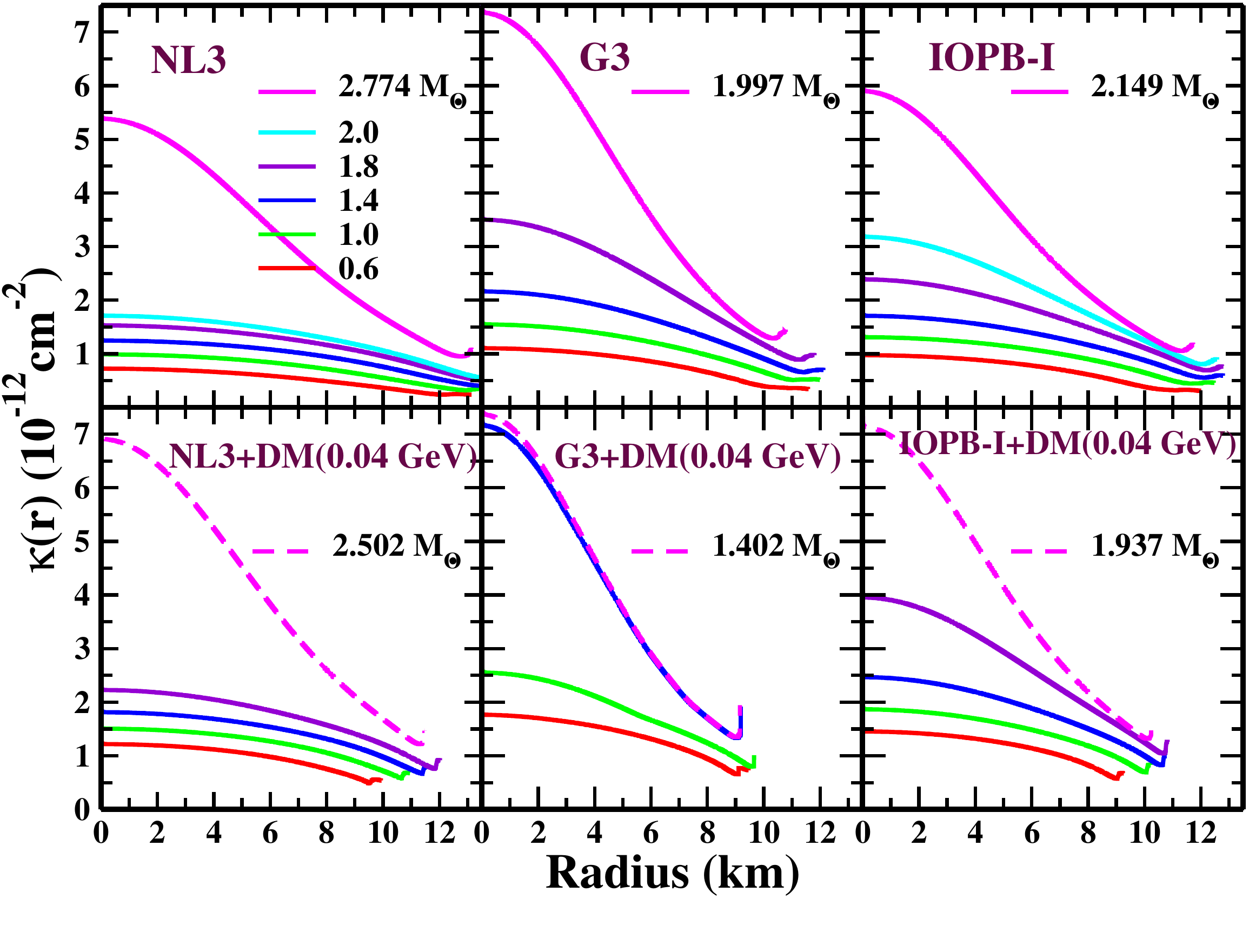}
        \caption{(colour online) The radial variation of ${\cal{K}}(r)$ without and with DM having momentum 0.04 GeV. The corresponding maximum mass is shown bold line (without DM) and dashed-line (with DM).}
        \label{curv2}
    \end{figure}

     Here, we calculate the ${\cal{K}}(r)$ within the NS. To see the parametric dependence of the curvature with radius, we fix the DM momentum at 0.04 GeV as shown in Fig. \ref{curv2} for different masses of the star. If one see carefully, the change of ${\cal{K}}(r)$ is not significant up to the canonical mass as compare to the maximum mass of the star. We calculate the change of ${\cal{K}}(r)$ with and without DM is $\approx$ 33 \%, and the change increases for maximum star mass. Hence, we inferred that the DM affects all the curvature parameters $\cal{K}$, $\cal{J}$, $\cal{R}$ and $\cal{W}$ of the NS considerably. 
    \begin{table}[tbp]
    \centering
    \caption{The central density ${\cal{E}}_c$, mass $M$, radius $R$, Surface curvature ${\cal{K}}(R)$, binding energy $B/M$ of the NS are tabulated with the variation of $k_f^{DM}$ both for canonical (1.4 $M_{\odot}$) and maximum mass star for NL3, G3 and IOPB-I parameter sets.}
    \label{table2}
    \renewcommand{\arraystretch}{1.5}
    \scalebox{0.7}{
    \begin{tabular}{ccccccccccccccccc}
    \hline\hline
    \multicolumn{1}{l}{\multirow{2}{*}{\begin{tabular}[c]{@{}l@{}}$k_f^{DM}$\\(GeV)\end{tabular}}} & \multicolumn{1}{l}{\multirow{2}{*}{\begin{tabular}[c]{@{}l@{}} \hspace{0.5cm}Star\\ \hspace{0.5cm}type\end{tabular}}} & \multicolumn{3}{l}{\begin{tabular}[c]{@{}l@{}}\hspace{0.7cm}${\cal{E}}_c$\\\hspace{0.3cm} (MeV fm$^{-3}$)\end{tabular}} &
    \multicolumn{3}{l}{\begin{tabular}[c]{@{}l@{}}\hspace{0.9cm}$M$\\\hspace{0.9cm}($M_\odot$)\end{tabular}}&
    \multicolumn{3}{l}{\begin{tabular}[c]{@{}l@{}}\hspace{0.9cm}$R$\\ \hspace{0.9cm}(km)\end{tabular}}& \multicolumn{3}{l}{\begin{tabular}[c]{@{}l@{}} \hspace{0.7cm}${\cal{K}}(R)$\\ \hspace{0.6cm}($10^{14}\cal{K}_{\odot}$)\end{tabular}}& 
    \multicolumn{3}{l}{\hspace{1.2cm}$B/M$}\\ 
    \cmidrule(lr){3-5}\cmidrule(lr){6-8}\cmidrule(lr){9-11} \cmidrule(lr){12-14}\cmidrule(lr){15-17}
    \multicolumn{1}{l}{} & \multicolumn{1}{l}{} &
    \multicolumn{1}{l}{NL3} & \multicolumn{1}{l}{G3} & \multicolumn{1}{l}{IOPB-I}&
    \multicolumn{1}{l}{\hspace{0.2cm}NL3} & \multicolumn{1}{l}{\hspace{0.2cm}G3} & \multicolumn{1}{l}{IOPB-I}&
    \multicolumn{1}{l}{\hspace{0.2cm}NL3} & \multicolumn{1}{l}{\hspace{0.2cm}G3} & \multicolumn{1}{l}{IOPB-I}&
    \multicolumn{1}{l}{\hspace{0.2cm}NL3} & \multicolumn{1}{l}{\hspace{0.2cm}G3} & \multicolumn{1}{l}{IOPB-I}&
    \multicolumn{1}{l}{\hspace{0.2cm}NL3} & \multicolumn{1}{l}{\hspace{0.2cm}G3} & \multicolumn{1}{l}{IOPB-I} \\ \hline
    \multicolumn{1}{l}{\multirow{2}{*}{}}0.00& \multicolumn{1}{l}{\hspace{0.5cm}Cano.}& 
    \multicolumn{1}{l}{270}& \multicolumn{1}{l}{460}   & \multicolumn{1}{l}{366}& 
    \multicolumn{1}{l}{1.400}    & \multicolumn{1}{l}{1.400}   & \multicolumn{1}{l}{1.400}&
    \multicolumn{1}{l}{14.08}    & \multicolumn{1}{l}{12.11}   & \multicolumn{1}{l}{12.78} &
    \multicolumn{1}{l}{1.477}    & \multicolumn{1}{l}{2.320}   & \multicolumn{1}{l}{1.977} & 
    \multicolumn{1}{l}{-0.084}    & \multicolumn{1}{l}{-0.098}   & \multicolumn{1}{l}{-0.092}\\ 
    \multicolumn{1}{l}{} & \multicolumn{1}{l}{\hspace{0.5cm}Max.} &            
    \multicolumn{1}{l}{870}& \multicolumn{1}{l}{1340}   & \multicolumn{1}{l}{1100} &
    \multicolumn{1}{l}{2.774}    & \multicolumn{1}{l}{1.997}   & \multicolumn{1}{l}{2.149} &
    \multicolumn{1}{l}{13.16}    & \multicolumn{1}{l}{10.78}   & \multicolumn{1}{l}{11.76}&
    \multicolumn{1}{l}{3.584}    & \multicolumn{1}{l}{4.695}   & \multicolumn{1}{l}{3.894}&
    \multicolumn{1}{l}{-0.207}    & \multicolumn{1}{l}{-0.162}   & \multicolumn{1}{l}{-0.165}\\ \hline
    \multicolumn{1}{l}{\multirow{2}{*}{}}0.02 & \multicolumn{1}{l}{\hspace{0.5cm}Cano.} &                        \multicolumn{1}{l}{286}    & \multicolumn{1}{l}{482}   & \multicolumn{1}{l}{385} &
    \multicolumn{1}{l}{1.400}    & \multicolumn{1}{l}{1.400}   & \multicolumn{1}{l}{1.400}&
    \multicolumn{1}{l}{13.63}    & \multicolumn{1}{l}{11.77}   & \multicolumn{1}{l}{12.42}&
    \multicolumn{1}{l}{1.626}    & \multicolumn{1}{l}{2.534}   & \multicolumn{1}{l}{2.153}&
    \multicolumn{1}{l}{-0.038}    & \multicolumn{1}{l}{-0.066}   & \multicolumn{1}{l}{-0.057}     \\ 
    \multicolumn{1}{l}{}  & \multicolumn{1}{l}{\hspace{0.5cm}Max.} &
    \multicolumn{1}{l}{890}    & \multicolumn{1}{l}{1400}   & \multicolumn{1}{l}{1120}&
    \multicolumn{1}{l}{2.734}   & \multicolumn{1}{l}{1.974}   & \multicolumn{1}{l}{2.118} &
    \multicolumn{1}{l}{12.91}    & \multicolumn{1}{l}{10.55}   & \multicolumn{1}{l}{11.54}&
    \multicolumn{1}{l}{3.741}    & \multicolumn{1}{l}{4.957}   & \multicolumn{1}{l}{4.061}&
    \multicolumn{1}{l}{-0.178}    & \multicolumn{1}{l}{-0.141}   & \multicolumn{1}{l}{-0.139}     \\ \hline
    \multicolumn{1}{l}{\multirow{2}{*}{}} 0.03 & \multicolumn{1}{l}{\hspace{0.5cm}Cano.} & 
    \multicolumn{1}{l}{320}    & \multicolumn{1}{l}{530}   & \multicolumn{1}{l}{430}& 
    \multicolumn{1}{l}{1.400}    & \multicolumn{1}{l}{1.400}   & \multicolumn{1}{l}{1.400} &
    \multicolumn{1}{l}{12.78}    & \multicolumn{1}{l}{11.09}   & \multicolumn{1}{l}{11.75}&
    \multicolumn{1}{l}{1.976}    & \multicolumn{1}{l}{3.034}   & \multicolumn{1}{l}{2.546}&
    \multicolumn{1}{l}{0.045}    & \multicolumn{1}{l}{-0.006}   & \multicolumn{1}{l}{0.016}     \\  
    \multicolumn{1}{l}{} & \multicolumn{1}{l}{\hspace{0.5cm}Max.} &                             
    \multicolumn{1}{l}{940}    & \multicolumn{1}{l}{1480}   & \multicolumn{1}{l}{1190} &
    \multicolumn{1}{l}{2.646}    & \multicolumn{1}{l}{1.923}   & \multicolumn{1}{l}{2.050}     &
    \multicolumn{1}{l}{12.39}    & \multicolumn{1}{l}{10.13}   & \multicolumn{1}{l}{11.06} &
    \multicolumn{1}{l}{4.097}    & \multicolumn{1}{l}{5.468}   & \multicolumn{1}{l}{4.492}&
    \multicolumn{1}{l}{-0.116}    & \multicolumn{1}{l}{-0.096}   & \multicolumn{1}{l}{-0.009}     \\ \hline
    \multicolumn{1}{l}{\multirow{2}{*}{}} 0.04 & \multicolumn{1}{l}{\hspace{0.5cm}Cano.}&                        \multicolumn{1}{l}{383}    & \multicolumn{1}{l}{640}   & \multicolumn{1}{l}{518}&
    \multicolumn{1}{l}{1.400}    & \multicolumn{1}{l}{1.400}   & \multicolumn{1}{l}{1.400}     &
    \multicolumn{1}{l}{11.60}    & \multicolumn{1}{l}{10.27}   & \multicolumn{1}{l}{10.76}&
    \multicolumn{1}{l}{2.638}    & \multicolumn{1}{l}{3.839}   & \multicolumn{1}{l}{3.317}&
    \multicolumn{1}{l}{0.016}    & \multicolumn{1}{l}{0.076}   & \multicolumn{1}{l}{0.105} \\  
    \multicolumn{1}{l}{} & \multicolumn{1}{l}{\hspace{0.5cm}Max.} &
    \multicolumn{1}{l}{1100}    &\multicolumn{1}{l}{1600}   & \multicolumn{1}{l}{1390}&
    \multicolumn{1}{l}{2.502}    & \multicolumn{1}{l}{1.837}   & \multicolumn{1}{l}{1.937} &
    \multicolumn{1}{l}{11.46}    & \multicolumn{1}{l}{9.53}   & \multicolumn{1}{l}{10.23}&
    \multicolumn{1}{l}{4.093}    & \multicolumn{1}{l}{6.274}   & \multicolumn{1}{l}{5.341}&
    \multicolumn{1}{l}{-0.023}    & \multicolumn{1}{l}{-0.026}   & \multicolumn{1}{l}{-0.002} \\ \hline \hline 
    \end{tabular}}
    \end{table}
    \begin{figure}[tbp]
        \centering
        \includegraphics[width=0.6\columnwidth]{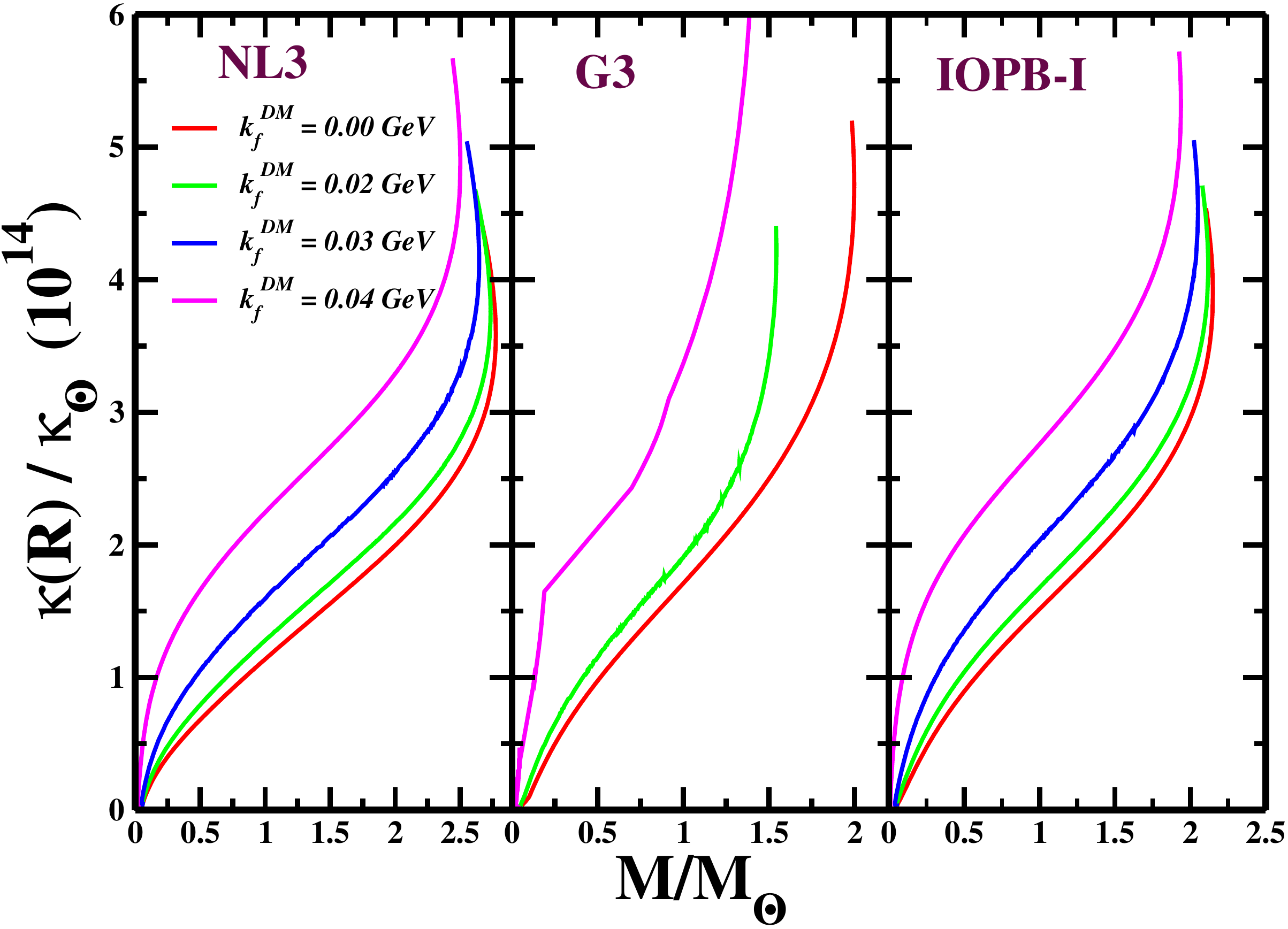}
        \caption{(colour online) The ratio of the surface curvature of NS and the Sun with the variation of NS mass with DM for NL3, G3 and IOPB-I.}
        \label{curv3}
    \end{figure}
   
    The curvature at the surface of the NS, which is more prominent to quantify the space-time wrap in the Universe. The variation of ${\cal{K}}(R)$/$\cal{K}_{\odot}$ with the mass of the NS is shown in Fig. \ref{curv3}. In our calculations, we find that the surface curvature of NS w.r.t the  Sun for NL3, G3 and IOPB-I parameter sets are 3.584, 4.695 and 3.894 respectively without including the DM and the values are 4.093, 6.478 and 5.341 with DM momentum (0.04 GeV). The surface curvature of the Sun  ($\cal{K}_{\odot}$) is $3.06\times 10^{-27}$~cm$^{-2}$  \citep{Kazim_2014}. The comparison of our results with the Sun gives the  ${\cal{K}}(R)$/$\cal{K}_{\odot}$ $\approx$ 10$^{14}$. The ratio ${\cal{K}}(R)$/$\cal{K}_{\odot}$ increases with the inclusion of DM. G3 parameter set provides softer EoS than IOPB-I, which indicates that the softer EoS facilitate us with larger surface curvature. The numerical values for both the canonical and maximum mass star are given in Table \ref{table2}. If the DM density is very high inside the NS, then the EoS becomes softer, which affects the curvatures significantly at the surface. More curvature means space-time is more curved and it depends on the amount of DM percentage inside the NS.
    
    From the above discussions, we led to conclude that the prediction of the strength of gravity within the NS is $\approx$ $10^{15}$ times more than the Sun. It increases few times with the addition of DM. In our calculations, the curvature $\cal K$ within the star decreases a few times towards the crust of the star and that order is $\approx$ 15 times than ${\cal{K}}_\odot$ without DM (shown in Fig. \ref{curv1}). The $\cal K$ increases with increase of DM density. The value of $\cal W$ is zero at the core and almost radially increases towards the crust. On the other hand, $\cal K$ is maximum at the core which comes mainly from the unconstrained parts of the NS. But the $\cal W$ has maximum value almost at the crust where GR play significant role. That means GR breaks down in the strong-field gravity while it retains its nobleness at the surface. From this analysis, one can say that GR is not well tested on the whole part of the NS than the EoS.
    
\begin{figure}[tbp]
         \centering
         \includegraphics[width=0.6\textwidth]{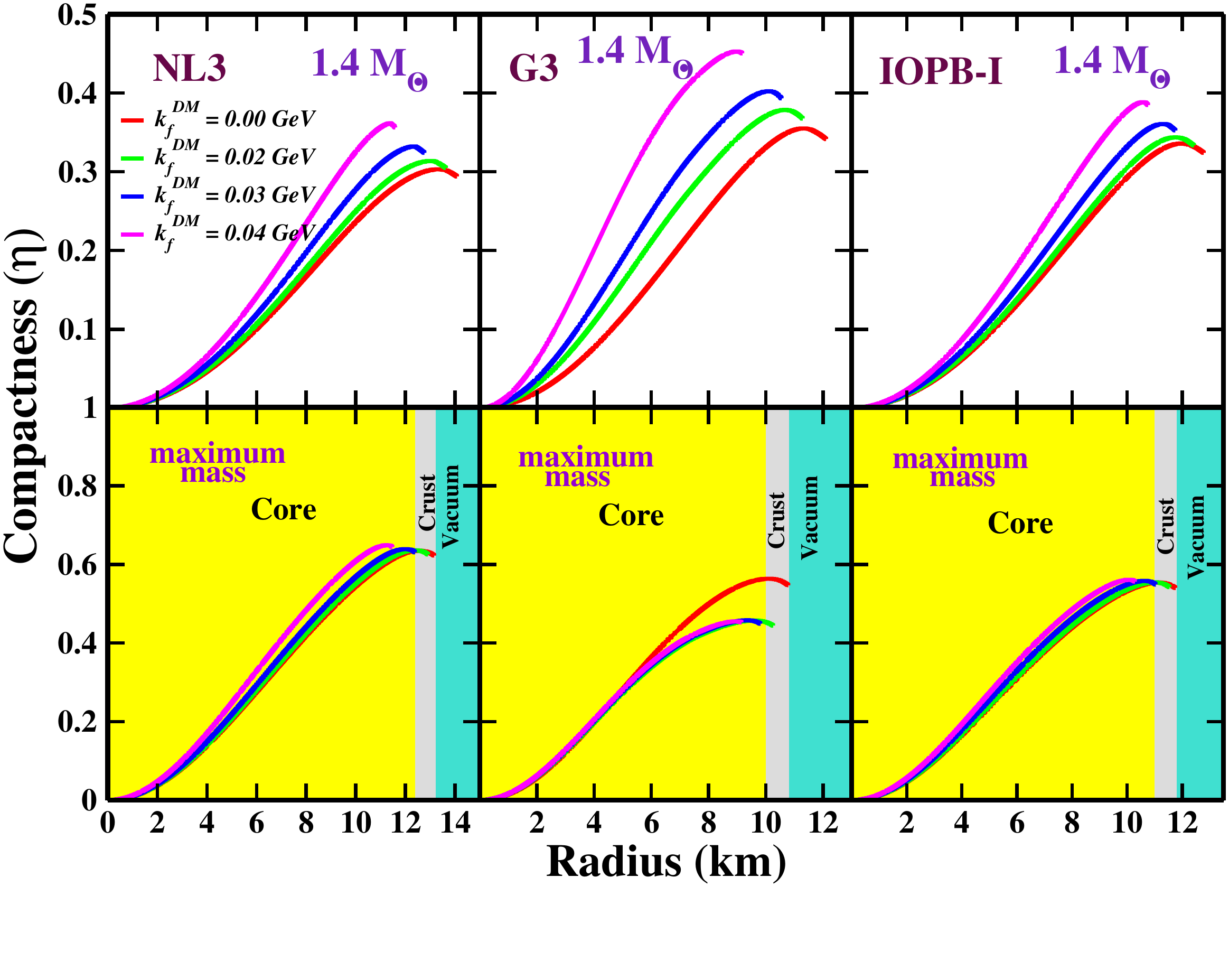}
         \caption{(colour online) The radial variation of compactness ($\eta$) for NL3 (left), G3 (middle) and IOPB-I (right) in the presence of DM.}
         \label{eta}
     \end{figure}

    The compactness ($\eta=\frac{GM}{r c^2}$) measures the degree of denseness of a star. The NS has larger mass and smaller radius as compared to the Sun, so its compactness is $10^5$ times larger than our Sun. Therefore to study the compactness of the NS, we plot the radial variation of the compactness of the NS in the presence of the DM, which is depicted in Fig. \ref{eta}. The compactness within the star increases with the increase of the DM momentum for different parameter sets, which is shown in Fig. \ref{eta}. It has a larger value for the maximum mass NS as compare to the canonical star.  With the increase of the DM percentage, the EoS becomes softer, which has less compactness as compared to the stiff EoS. The compactness is maximum at the surface of the star.
    
    \subsection{Binding energy of the NS}
    The gravitational binding energy ($B$) is defined as the difference between the gravitational mass ($M$) and  baryonic mass ($M_B$) of the NS;  $B=M-M_B$, where $M$ is calculated as \citep{NKGb_1997, Xiao_2015}
    \begin{equation}
    M=\int_{0}^{R} dr \  4\pi r^2 {\cal E}(r),
    \end{equation}
    and $M_B=Nm_b$, where $m_b$ is the mass of baryons (931.5 MeV) and $N$ is the number of baryons calculated by integrating over the whole volume in the Schwarzchild limit as 
    \begin{eqnarray}
    N=\int_{0}^{R} dr \ 4\pi r^2 \Big[1-\frac{2m(r)}{r}\Big]^{-1/2}.
    \end{eqnarray}
    The $N$ is found to be $\approx10^{57}$ same as given in the Ref. \cite{NKGb_1997}. In our work binding energy $B$ originally corresponds to $B/M$, which is more convenient for comparison purpose. 
    The binding energy per particle of the symmetric NM is $\approx$ -16 MeV, i.e. it needs 16 MeV to make the system unbound. For pure neutron matter (PNM) system, it is positive \citep{Serot_1986}. That means the PNM system is already unstable. It is well acknowledged that the nuclear force is state dependent and the nucleon-nucleon interaction is divided into three categories- singlet-singlet, triplet-triplet and singlet-triplet. The singlet-singlet and the triplet-triplet interaction are repulsive in nature while the singlet-triplet interaction is attractive \citep{Patra_1992,Satpathy_2004,Kaur_2020}. Due to the excess number of neutrons, the repulsive part adds instability to NS. However, its enormous gravitational force balances the repulsive nuclear force. Thus for the whole NS, the $B$ is negative.
    
\begin{figure}
        \centering
        \includegraphics[width=0.6\columnwidth]{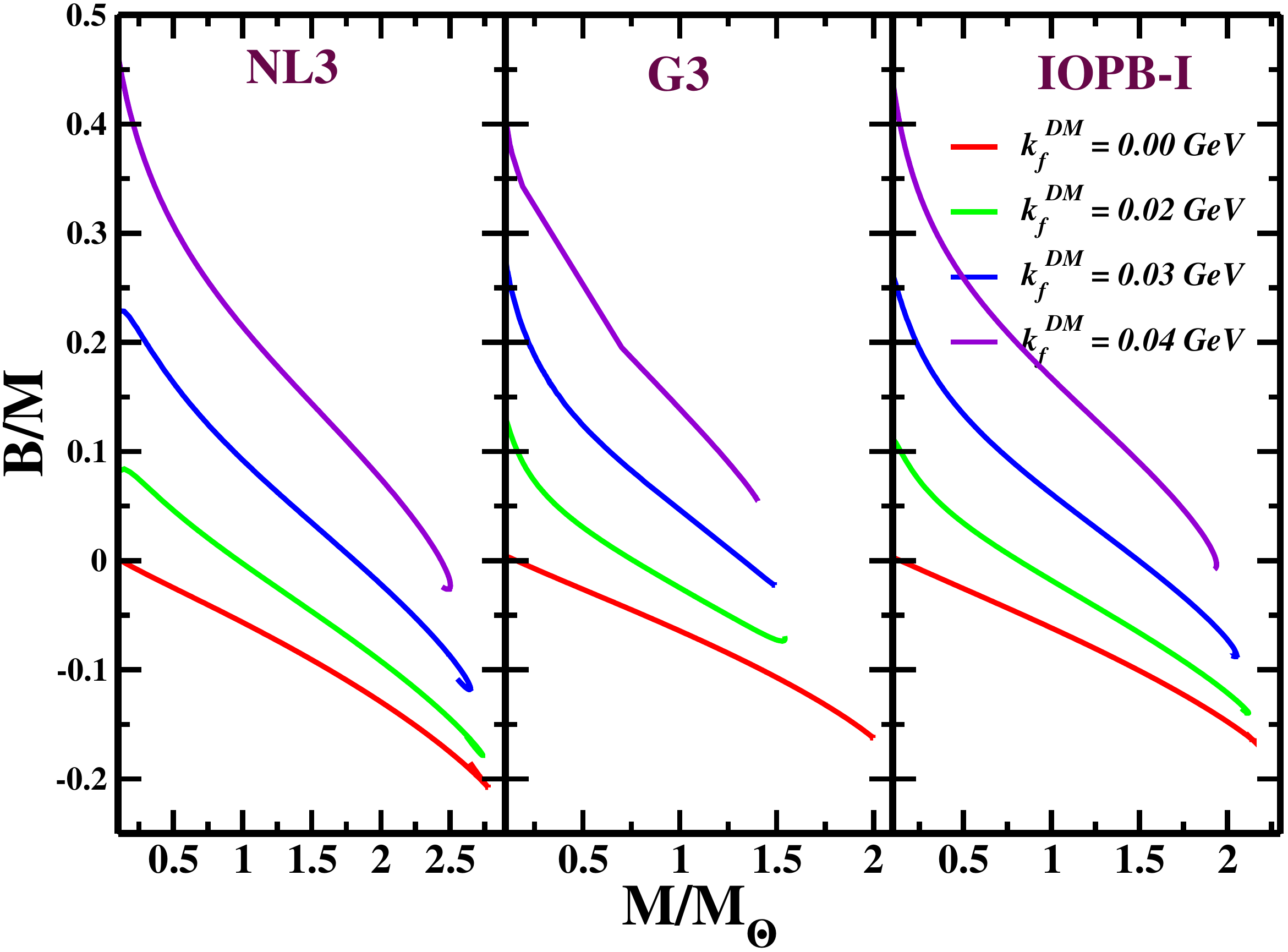}
        \caption{(colour online) The variation of $B/M$ with the M/$M_{\odot}$ of the NS with and without DM.}
        \label{BE}
    \end{figure}

    With the addition of DM inside the NS, the $B$ step up towards positive, that means it is going to be unstable. However, the instability of NS depends on the DM percentage. The variation of $B/M$ with $k_f^{DM}$ is depicted in Fig. \ref{BE} for assumed parameter sets. The numerical values given in Table \ref{table2}. A careful inspection of the Table \ref{table2} shows that up to 0.02 GeV the $B/M$ of the canonical and maximum mass NS are negative. It indicates that both the canonical and maximum mass NS are bound systems with this amount of DM. However, if we increase the DM momentum, the canonical NS system becomes unbound with positive $B$. For example, with DM momentum 0.04 GeV the binding energy for the canonical star becomes positive for different parameter sets. But still, the maximum mass NS shows a bound system with negative binding energy. From this, we conclude that one can constraint the DM percentage inside the NS. If the DM contained is more than the canonical star, it forms a mini black hole at the core and destroys the NS \citep{Goldman_1989, De_Lavallaz_2010, Kouvaris_2011, Kouvaris_2012}. The cooling of NS is also faster with the increasing of DM mass \citep{Ding_2019, Bhat_2019}. That means the positive $B$ may have a relation with the cooling properties of the NS or in other words; it may accelerate the Urca process. 

    \section{Conclusion}
    \label{con}
     In the present work, we study the impacts of DM on the curvatures of the NS. The calculations are done with some well tested RMF parameters sets like NL3, the E-RMF parameter IOPB-I (two additional couplings to RMF) and G3 parameter (six extra couplings to RMF). The E-RMF formalism well suited to both finite nuclei properties as well as NM in both normal and extreme conditions. The parameter sets G3 and IOPB-I  establish the recently measured the maximum mass and the radius of the NS. The EoSs of the NS are calculated by assuming that the DM particles inside the star. We find that the DM plays a significant role in the NS curvatures, even in the E-RMF model, which yield a softer EoS.
    
     We calculate various curvatures with the variations of the baryon density, mass and radius of the NS in the presence of the DM. The curvature increases or decreases with the baryon density. It is observed that at lower density the quantities $\cal{K}$ and $\cal{W}$ gives more curvature than $\cal{J}$, $\cal{R}$. At the crust region, the curvatures $\cal{J}$, and $\cal{R}$ almost vanish. The $\cal{K}$ and $\cal{W}$ approach each other within the crust and have a local maximum in that region. Moreover, the radial variation of ${\cal{K}}(r)$ increases with the increasing of DM momentum for maximum mass and it has very small effects up to canonical star. The percental change of ${\cal{K}}(r)$ with and without DM is approximately 33\% for 1.4 $M_{\odot}$, and it increases for the maximum mass. From the surface curvature study, we conclude that the softer EoS gives large curvature than the stiffer one. Both curvature ($\cal W$) and compactness ($\eta$) approach each other at the crust. Assuming the EoS of the crust is not well defined and hence the crust is the best site to measure the deviation of GR in the strong-field gravity limit, activity of pulsar glitch etc. The binding energy increases towards positive with the increase of DM momentum which makes the NS unstable. From this, we conclude that a tiny amount of DM can accumulate inside the NS. The more percentage of the DM heat the NS, and it accelerates the Urca process, which enhance the cooling of the NS, and it makes the NS unstable.
     \section*{Acknowledgement}
     SKB is supported by the National Natural Science Foundation of China Grant No. 11873040\\
     
\bibliographystyle{JHEP}
\bibliography{curvature.bib}
\end{document}